\documentclass[10pt, letterpaper,twocolumn]{article}
\usepackage{amsfonts}
\usepackage{graphics}
\usepackage{amsmath}
\usepackage{authblk}
\usepackage{hyperref}

\hypersetup{breaklinks=true}

\usepackage{graphicx} 
\usepackage{algorithm}
\usepackage[sorting=none]{biblatex}
\usepackage{algpseudocode}
\usepackage[left=1.5cm, right=1.5cm, top=1.5cm, bottom=1.5cm]{geometry}

\usepackage{amssymb}
\title{\fontsize{14}{14}\selectfont  A Surprisingly Simple Method for Distributed Euclidean-Minimum Spanning Tree / Single Linkage Dendrogram Construction from High Dimensional Embeddings via Distance Decomposition}

\author[1]{Richard Lettich}

\affil[1]{
\fontsize{9}{11}\selectfont
Lawrence Berkeley National Laboratory
}

\date{}
\begin{document}


\twocolumn[
\begin{@twocolumnfalse}
\vspace{-3em}
\maketitle
\vspace{-4em}
\begin{abstract}
We introduce a communication-aware decomposition method for the distributed calculation of exact Euclidean Minimum Spanning Trees in high dimensions (where sub-quadratic algorithms are not effective), or more generalized geometric-minimum spanning trees of complete graphs, where for each vertex \(v\in V \) in the graph \(G=(V,E)\) is represented by a vector in \(\vec{v}\in \mathbb{R}^n\), and each for any edge, the the weight of the edge in the graph is given by a symmetric binary `distance' function between the representative vectors \(w(\{x,y\}) = d(\vec{x},\vec{y})\). This is motivated by the task of clustering high dimensional embeddings produced by neural networks, where low-dimensional algorithms are ineffective; such geometric-minimum spanning trees find applications as a subroutine in the construction of single linkage dendrograms, as the two structures can be converted between each other efficiently.
\end{abstract}
\end{@twocolumnfalse}
]
\paragraph{Algorithm}
This algorithm generates the minimum spanning tree for the complete graph of points. It utilizes a ``dense'' minimum spanning tree subkernel which operates on the vectors, which we'll denote d-MST, which forms the MST from vectors that describe the vertices’ positions and the definition of the distance function (most likely via all pairs brute-force), alongside a more typical Minimum Spanning Tree Algorithm (MST) which operates on sparse graphs.  This two layer definition allows us to efficiently parallelize the construction of high dimensional Euclidean-Minimum Spanning Tree and similar graphs with minimal communication, and exploit existing high performance kernels for E-MST (or d-MST) calculation without adjustment to enable distributed parallelism.

The algorithm operates by performing a number of calls to the d-MST algorithim using subgraphs of the original graph, or concretely, subsets of the vectors being operated on. The union of the d-MSTs of these subgraphs forms a superset of the MST of the entire graph. Taking the MST of this union, which involves only \(O(|V|\times |P|)\) edges (where \(|P|\) is the number of sets in our partition) yields the MST of the complete graph, and is a relatively inexpensive operation.

For conceptual clarity, we assume that the indexing scheme is automatically maintained. However, in actual implementation, reindexing the vertices that describe the MST subgraphs to respect all vertices of the original graph would be necessary to respect the global vector indexinging upon return of each d-MST.

\begin{algorithm}[]
\caption{Decomposed Euclidean-Minimum Spanning Tree Algorithm}\label{alg:cap}
\begin{algorithmic}
\State \(P = \{S_i\}_{i\in \{1..n\}} \gets\) Partition of Vectors (Vertices), \(V\)\;
\State TreeEdges \(\gets \emptyset\)\;
\For{ \(j \in \{2..n\} \)}
\For{ \(i \in \{ 1..j-1\}\) }
   \State  TreeEdges \(\gets\)  TreeEdges \(\cup\) d-MST(\(S_i\cup  S_j\)) \;
\EndFor
\EndFor
\State TreeEdges \(\gets\)  MST(TreeEdges)
\end{algorithmic}
\end{algorithm}

We give the sequential pseudocode algorithm in \ref{alg:cap}, which trivially admits parallelization to \(\frac{|P|(|P|-1)}{2}\) processes, assuming each process has access to a d-MST kernel,  by giving each process  a  subgraph d-MST to compute, performing a Gather and unioning  the resulting trees, thereafter taking the MST of the resultant graph.

We present (a forest generalized) proof of correctness of algorithm \ref{alg:cap} in theorem \ref{MSFTheorem}, which depends on lemma 1. Note, we assume the minimum spanning forest is unique.

Intuitively, Lemma 1 gives an optimal substructure property for MSTs, which tells us that for any subgraph defined by only considering some vertices of a graph, the MST of said subgraph contains every edge in the MST of the whole graph which involves only those vertices. In Theorem 1 we give the proof of Algorithm 1 by decomposing the graph into sub-graphs whose union is the complete graph, and take the MST of each subgraph. By the optimal substructure property of Lemma 1, the union of the MSTs of these subgraphs is a superset of the MST of the entire graph.

\newtheorem{lemma}{Lemma}
\begin{lemma}
Let \(G=(V,E)\) be a graph, \(S\subseteq V\) be any subset of vertices, and \(G[S]\) denote the induced subgraph. For all \(G\) and \(S\), the MSF obeys an optimal substructure property:
\[\text{MSF}(G)[S] \subseteq \text{MSF}(G[S]).\]
\end{lemma}
Suppose there exists some edge $e\in \text{MSF}(G)[S]$. Consider the connected component $T=(V_T,E_T)$ of the original graph which it is a part of. Choose any size-two partition $V_T=Q\cup Q^c$ of $V_T$ such that the two vertices incident to $e$ are in separate sets. By the cut property, and by being part of $\text{MSF}(G)$, we have that $e$ is the minimal-weight edge in $T$ with one incident vertex in $Q$ and one in $Q^c$. Additionally, because it occurs in $S$, it must be the minimal-weight edge in $T[S]$ with one vertex incident to $Q\cap S$ and one to $Q^c\cap S$, implying it must be in $\text{MSF}(T[S])$, and therefore $\text{MSF}(G[S])$. \hfill \(\blacksquare\)
\newtheorem{T}{Theorem}

\begin{samepage}
\begin{T}\label{MSFTheorem}
Let \(G=(V,E)\) be a symmetric graph, and \(P=\{S_i\}_{i\in\{1..n\}}\) be a partition of V. Then:
\[MSF(G) = MSF\big(\bigcup_{1 \leq i <  j \leq n} MSF(G[S_i\cup S_j])\big).\]
\end{T}
\scalebox{0.75}{
\begin{minipage}{\linewidth}
\begin{align*}
    MSF(G) &= MSF(G)\cap V\times V & V\times V \text{is universe set} \\
      &= MSF(G)\cap \bigcup_{S_i,S_j \in P \times P} S_i\times S_j& \text{Def of partition}\\
      &= \bigcup_{S_i,S_j \in P \times P} MSF(G) \cap(S_i\times S_j) &\text{Dist of intersect over union}\\\
      &= \bigcup_{1 \leq i< j \leq n } MSF(G) [S_i \cup S_j] &\text{\small symmetry; Def induced subgraph} \\
         &\subseteq \bigcup_{1 \leq i< j \leq n } MSF(G [S_i \cup S_j]) & \text{Lemma 1}\\
\end{align*}
\newcommand{\qed}{\tag*{$\blacksquare$}}
\end{minipage}
}
\end{samepage}
We then consider to the two subset relations:
\[ MSF(G)\subseteq  \bigcup_{1 \leq i< j \leq n }  MSF(G [S_i \cup S_j])  \qquad \subseteq G.\]
By combining these relations with the cut property, we can infer
\[MSF(G) = MSF\big(\bigcup_{1 \leq i <  j \leq n} MSF(G[S_i\cup S_j])\big).\tag*{$\blacksquare$}\]
\paragraph*{Cost Analysis}

The Bandwidth required for the final gather is \(O(|V||P|)\). Recall \(p=\frac{|P|(|P|-1)}{2}\) is our number of processors. Thus in terms of number of vectors \(|V|\) and processors \(p\), our cost is \(O(|V|\sqrt{p})\). While contrived, this could be instead replaced by a reduction operation on the trees produced, where the binary operation is given by \(\oplus(T_1,T_2) = MST(T_1 \cup T_2)\) reducing the bandwidth cost to \(O(|V|)\),  but we find this purely pedantic.

Let \(f\) denote the work to performed by the d-\(MST\) kernel. Then we observe the algorithm, given \(|P|\) evenly sized partitions, performs \[\frac{|P|(|P|-1)}{2} f(2\frac{|V|}{|P|})\]
work. If \(f \in \Omega(|V|^2)\), we are performing up to \(\lim_{P\to \infty}\frac{2 (|P|-1)}{|P|} = 2\) times as much work in terms of operations performed by the d-MST kernel as compared to not decomposing the calculation.

This redundancy arises from the algorithm's structure, where the double loop over the partitions leads to repeated calculations of edges within each subset \(S_i\) and \(S_j\). Specifically, for each call to d-MST\((S_i \cup S_j)\), we calculate all edges in \((S_i \cup S_j) \times (S_i \cup S_j)\), which includes the edges in \(S_i \times S_i\) and \(S_j \times S_j\). As a result, the edges within each subset are computed multiple times across different iterations of both the outer and inner loops (i.e., for various \(S_j\) and \(S_i\)).
\paragraph*{Related Work}
This work was inspired by \textit{Driscoll et al.}'s paper on communication-optimal algorithms for \(N\)-Body simulations~\cite{Driscoll}, and preceding work by \textit{Plimpton}~\cite{PLIMPTON19951}, \textit{Solomonik and Demmel}~\cite{Solomonik2011}, and \textit{Ballard et al.}~\cite{Ballard_2011}.

\textit{Wang et al.}~\cite{Wang2021} give a shared-memory method for producing E-MST / Single Linkage Dendrograms in low dimensions.

NVIDIA~\cite{raschka2020machine} has produced a single GPU-accelerated method for the construction of E-MST / Single Linkage Dendrograms in high dimensions based on the work of \textit{Arefin et al.}~\cite{Arefin2012}.

\printbibliography

\end{document}